\documentclass[12pt,preprint,psfig]{aastex}
\newcommand{\etal}{{et al.} }
\newcommand{\rxte}{{\it RXTE} }

\newcommand{\asca}{{\it ASCA} }

\newcommand{\xmm}{{\it XMM-Newton} }
\newcommand{\xmmp}{{\it XMM-Newton}}
\newcommand{\chandra}{{\it Chandra} }

\newcommand{\suzaku}{{\it Suzaku} }
\newcommand{\suzakup}{{\it Suzaku}}
\newcommand{\bepposax}{{\it BeppoSAX }}

\newcommand{\feka}{{Fe~K$\alpha$} }

\newcommand{\fekalfa}{{Fe~K$\alpha$} }

\newcommand{\fekbeta}{{Fe~K$\beta$} }

\newcommand{\feklya}{{Fe~{\sc xxvi}~Ly$\alpha$} }

\newcommand{\bsax}{{\it BeppoSAX} }

\newcommand{\figplrat}{{Fig.~1} }

\newcommand{\figcxmoratp}{{Fig.~2}}
\newcommand{\figcxmorata}{{Fig.~2a} }

\newcommand{\figcxmoratb}{{Fig.~2b} }

\newcommand{\figcxmoratc}{{Fig.~2c} }

\newcommand{\figbestmodel}{{Fig.~3} }

\newcommand{\figrgsoverlay}{{Fig.~4} }

\newcommand{\figidvsin}{{Fig.~5} }
\newcommand{\figidvsinp}{{Fig.~5}}
\newcommand{\figky}{{Fig.~6} }

\newcommand{\figkya}{{Fig.~6(a)} }

\newcommand{\figkyb}{{Fig.~6(b)} }

\newcommand{\tablespfit}{Table~1 }
\newcommand{\tablespfitp}{Table~1}
\newcommand{\tablehistfit}{Table~2 }
\newcommand{\tablehistfitp}{Table~2}
\newcommand{\tablekymodel}{Table~3 }
\newcommand{\tablekymodelp}{Table~3}

\newcommand{\ergs}{${\rm erg \ cm^{-2} \ s^{-1}}$ }
\newcommand{\ergsp}{${\rm erg \ cm^{-2} \ s^{-1}}$}

\usepackage[dvips]{color}

\begin{document}
\title{NGC~2992 in an X-ray high state observed by \xmmp: Response of
the Relativistic \feka Line to the Continuum}
\author{X. W. Shu\altaffilmark{1,2}, T. Yaqoob\altaffilmark{2,3}, 
K. D. Murphy\altaffilmark{2,4}, V. Braito\altaffilmark{5}, 
J. X. Wang\altaffilmark{1,6}, W. Zheng\altaffilmark{2} }
\altaffiltext{1}{Center for Astrophysics,
University of Science and Technology of China, Hefei, Anhui 230026, China. 
xwshu@mail.ustc.edu.cn}
\altaffiltext{2}{Department of Physics and Astronomy,
Johns Hopkins University, 3400 N. Charles Street., Baltimore, MD 21218, USA.
yaqoob@skysrv.pha.jhu.edu}
\altaffiltext{3}{Astrophysics Science Division,
NASA/Goddard Space Flight Center, Greenbelt, MD 20771, USA.}
\altaffiltext{4}{Massachusetts Institute of Technology, Center for
Space Research, NE80$-$6013, 77 Massachusetts Avenue,
Cambridge, MA~02139, USA.}
\altaffiltext{5}{X-ray Astronomy Group, Department of Physics and
Astronomy, Leicester University, Leicester, LE1 7RH, UK.}
\altaffiltext{6}{Key Laboratory for Research in Galaxies and Cosmology, University of Sciences and Technology of China, Chinese Academy of
Sciences}

\begin{abstract} 
We present the analysis of an
\xmm observation of the Seyfert galaxy NGC~2992. 
The source was found in its highest level of X-ray activity yet detected,
a factor $\sim 23.5$ higher in 2--10~keV flux than the historical
minimum. NGC~2992 is known to exhibit X-ray flaring activity
on timescales of days to weeks, and the \xmm data provide at
least factor of $\sim 3$ better spectral
resolution in the Fe~K band than any previously measured
flaring X-ray state. We find that there is a broad feature
in the $\sim 5-7$~keV band which could be interpreted as a relativistic
\fekalfa emission line. Its flux appears to have increased
in tandem with the 2--10~keV continuum when compared
to a previous \suzaku observation when the continuum was
a factor of $\sim 8$ lower than that during the \xmm observation.
The \xmm data are consistent with the general picture that
increased X-ray activity and corresponding changes in the
\fekalfa line emission occur in the innermost regions of the putative
accretion disk. This behavior contrasts with the
behavior of other AGN in which the 
\fekalfa line does not respond to
variability in the X-ray.

\end{abstract}


\section{Introduction}
\label{intro}

Relativistically broadened \fekalfa emission lines in 
active galactic nuclei (AGNs) and Galactic Black-Hole Binaries
(GBHB) are potentially a powerful probe of 
accreting black-hole systems. In principle, modeling the 
\fekalfa line
profile, supplemented with assumptions about its radial
and angular emissivity function in the accretion disk, can yield
information on the inclination angle of the disk, its radial
extent, the proximity of its inner edge 
to the black-hole, and the angular
momentum, or spin of the black hole 
(e.g. Dov\v{c}iak \etal 2004a; Beckwith \& Done 2004;
Brenneman \& Reynolds 2006).
Whilst there are plenty of studies in the literature that
address the {\it occurrence} and modeling of the time-averaged
broad \fekalfa line profiles (e.g. Guainazzi \etal 2006;
Nandra \etal 2007; Miller \etal 2007, and references therein), 
observational data on the {\it variability} of the relativistic
lines in AGNs is remarkably sparse 
(e.g. see Fabian 2006, and references therein). Variability of
the line shape and intensity, especially in relation to
the variability of the continuum, would provide stronger
and more model-independent information on the accretion-disk
and black-hole parameters (in particular on the unknown radial emissivity
function). It would also lay the groundwork for 
future reverberation mapping
of the black-hole metric in the strong gravity regime 
(e.g. Reynolds \etal 1999).

However, there are only two AGNs for which it has
been possible to assess the variability of the broad \fekalfa
line and in both of those cases it is found that the
line intensity does not respond to the observed X-ray
continuum (MCG~$-$6-30-15: e.g. Fabian \etal 2002; NGC~4051: Ponti \etal 2006a).
{Strictly speaking the NGC~4051 results refer
to the reflection continuum appearing to 
correlate with the intrinsic continuum at small
fluxes, but then remaining roughly constant
as the intrinsic continuum flux increases. However, this interpretation
is model-dependent because the intrinsic continuum shape and
flux is determined indirectly.}
This behavior has been interpreted in terms of the
so-called ``light-bending'' model in which physical motion
of the X-ray source in the strong gravity field of the
black hole causes variability in the X-ray continuum for an
observer ``at infinity'' but the illumination of the disk
(and therefore the \fekalfa line flux) remains relatively
unaffected. 
It is the variable amplification effect of light-bending
that causes the X-ray continuum variability--any intrinsic
variability of the luminosity of the X-ray source must
be suppressed or absent, otherwise the \fekalfa line
intensity would once again vary.
In a few other AGN, although the variability of
the \fekalfa line cannot be isolated, the overall
behavior of the disk-reflection and fluorescence spectrum 
(of which the emission
line is a part), has been interpreted as being consistent
with this scenario (e.g. 1H~0707$-$495, Fabian \etal 2004;
1H~0419$-$577, Fabian \etal 2005; IRAS~13224$-$3809, Ponti \etal 2006b).
Similar non-variability of the relativistic \fekalfa line
has been observed in a GBHB (XTE~J16550--Rossi \etal 2005), and this
has also been interpreted in the context of the light-bending 
model. The origin of the physical motion of the
X-ray source (relative to the black-hole) and it's geometry
remain unspecified in the light-bending model.
We note that alternative interpretations of the apparent
broad \fekalfa emission lines in the above-mentioned
(and other) sources in terms of absorption-only models remain viable
and are still a matter of considerable debate (e.g. Miller \etal 2008, 2009;
Turner \etal 2009; Turner \& Miller 2009 and references therein).

The reason why so little observational data exists on
the variability of the relativistic \fekalfa line, in AGN at least,
is because several criteria must be met by the source at once,
and such objects are not common. Firstly,
the current sensitivity of X-ray detectors require that the source
is bright enough that the time-sliced spectra 
have sufficient signal-to-noise.
Secondly, there must be be a sufficient quantity of data 
(commensurate with the timescale
of variability) in order to meaningfully test variability models.
Thirdly, the source must have a well-measured broad \fekalfa line with a
large enough equivalent width that
its contrast against the continuum is high 
enough to constrain the line profile parameters 
with sufficient accuracy. Fourth, 
the amplitude of variability of the {\it continuum} should
be large enough in order to be able to test whether the 
{\it line} flux varies in response to it.
Although there are plenty of AGN that satisfy one of these
criteria, sources that satisfy all four are very rare.
Indeed MCG~$-$6-30-15 and NGC~4051 are the only two AGN
for which a definitive result has been reported
on the non-response of
the broad \fekalfa to continuum variations (as opposed
to response of the broadband reflection spectrum).
In the present paper we show that 
non-variability of the relativistic \fekalfa 
line in NGC~2992 in response to an 
increase in the X-ray continuum level is ruled out, in contrast to the known
behavior of
the other AGN discussed above.

NGC~2992 is a nearby ($z=0.00771$) Seyfert galaxy that
has been observed by every major X-ray mission over the past $\sim30$ years,
and has shown hard X-ray flux variability by more than a factor 
of $20$ (Piccinotti \etal 1982; Turner \& Pounds 1989; Turner \etal 1991;
Nandra \& Pounds 1994; Weaver \etal 1996; Gilli \etal 2000;
Matt \etal 2003; Beckmann \etal 2007; Murphy \etal 2007;
Yaqoob \etal 2007). 
In the present paper we describe the results of an \xmm observation of 
NGC~2992 made in 2003, May. Some results from this observation
from spectral fitting over a restricted bandpass ($2.5-10$~keV) have
been presented by Brenneman \& Reynolds (2009). However,
Brenneman \& Reynolds (2009) did not 
account for the heavy photon pile-up in the \xmm CCD data,
which considerably distorts the spectrum.
Here we examine the data in detail, over the full \xmm bandpass, and in the
context of the historical X-ray behavior of NGC~2992.
During the \xmm observation NGC~2992
was at the top end of its historical range of 
activity in terms of the 2--10~keV flux.
We detected a strong, broad emission-line
feature in the Fe~K band which implies extreme
variability in line flux compared to historical \suzaku data.
This paper is organized as follows.
In \S\ref{obs} we describe the \xmm observation and data reduction. 
Spectral analysis of the data is described in \S\ref{fitting}.
In \S\ref{results} we discuss the results in detail
and in \S\ref{historicaldata} we compare
the \xmm results with those from historical data.
In \S\ref{conclusions} we discuss some general implications
of our findings and present our conclusions.
A cosmology with $H_0 = 70$ km s$^{-1}$ Mpc$^{-1}$, $\Lambda=0.73$, $\Omega=1$ 
is assumed throughout.

\section{Observation \& Data Reduction}
\label{obs}

NGC 2992 was observed by \xmm on 2003 May 19, for a duration 
about 29 ks. The EPIC pn camera and the two MOS cameras were operated in 
Full Window mode with a medium thick filter.
We used principally the pn data, which have much higher sensitivity, using 
the MOS data only to check for consistency. The RGS spectral bandpass
does not cover the Fe~K region and the signal-to-noise is poor,
but we will show that our model for the pn data is consistent with the
RGS spectra (see \S\ref{baselinemodel}).
The calibrated event lists were extracted 
from the observation data files using the EPCHAIN pipeline tasks provided by 
the \xmm Science Analysis System (SAS) version 7.0.0, using the 
latest calibration files available at the time of the analysis
(February 2008).
Spectra and light curves for both the source and background were
extracted using
XSELECT and FTOOLS (version 6.3.2).
Data were selected 
using different event patterns in order
to investigate the effects of pile-up.
Specifically, we extracted single-pixel events for both
the pn and MOS,
as well as single plus double-pixel events
(patterns 0--4) for the pn, and patterns 0--12 for the MOS.
We first extracted source spectra from 40'' radius circles centred on the 
source position and assessed
the extent of pile-up using the SAS task $epatplot$. 
We found that all the spectra from
both \xmm EPIC pn and MOS data at the core of the PSF for NGC 2992 were 
affected by heavy photon pile-up.
Therefore, 
using the $epatplot$ tool we compared the spectra extracted in a number of 
annuli with the expected distribution of energy and event grades with radius. 
We found that the smallest 
acceptable inner radius for the annular extraction region
was 10''. Moreover, we found that the pn spectra made from the
different pattern selections were consistent with each other.
Therefore, we extracted source counts in annuli between 
10''-40'', selecting for patterns 0--4 (pn) and 0--12 (MOS), since
these event-pattern selections yielded spectra with higher
signal-to-noise ratios than spectra extracted from single-pixel events.

For both pn and MOS data,
the background events were extracted from source-free areas
on the same CCD using two rectangular regions with a combined area 4.5 
times larger than the source region. Background flares were present at the 
end of the observation.  We adopted a count-rate filtering criterion 
from the full-band ($\sim 0.3-12$~keV) lightcurves
of 0.2 ct s$^{-1}$ (pn) and 0.1 ct s$^{-1}$
(MOS) to remove the high particle background, which resulted 
in a net exposure time of $\sim$ 21.4~ks for the pn
and 24.3~ks and 24.0~ks for MOS1 and MOS2 respectively.
The response matrix and ancillary response file for the pn spectrum were 
generated using the RMFGEN and ARFGEN tools within the SAS software. 
We will show in \S\ref{prelimfit} 
that the broadband 0.5-10 keV pn spectrum agrees well with the summed MOS1 and 
MOS2 spectra, indicating that we successfully
mitigated the effect of pile-up since the severity of effect of pile-up is
different for the pn and MOS data.
We will further show, in \S\ref{baselinemodel}, that the 
pn and RGS data are consistent with each other,
indicating that any remaining
effects of pileup are negligible in the overlapping energy band.

\section{Spectral Analysis}
\label{fitting}

\subsection{Preliminary Spectral Fitting} 
\label{prelimfit}

During the \xmm observation, the source did not vary significantly in flux
so we do not show the lightcurve. 
The 
excess variance ($\sigma^{2}_{\rm rms}$, see e.g. Turner et al. 1999;
Markowitz \& Edelson 2004) 
from a 0.5--10~keV pn light curve binned at 256~s 
(using only fully exposed bins) was
found to be $\sigma^{2}_{\rm rms}$ = 0.00078$\pm$0.00018.
This value is at the lower end of the range
that is typical for AGN with an
intrinsic 2--10 keV luminosity similar to that of NGC~2992 
(see \S\ref{contresults}).
For comparison, the comparable excess variance for one of the
most highly variable low-luminosity AGN, NGC~4051, can be as high
as $0.162 \pm 0.0243$ (Turner \etal 1999). 
Therefore, the spectral analysis described 
in the following was performed on the time-averaged spectrum. 

We performed spectral fitting 
to the NGC~2992 \xmm EPIC pn data in the 0.5--10 keV range using XSPEC 
version 11.3.2 (Arnaud 1996). The spectrum was binned 
uniformly to $\sim 40$ eV per bin and we verified
that this resulted in $>25$ counts per bin across
the full energy range, validating the use of $\chi^{2}$ 
minimization for spectral fitting. All statistical errors given
hereafter correspond to 
90\% confidence for one interesting parameter ($\Delta \chi^2=2.076$), 
unless stated otherwise. In all of the model fitting, the 
Galactic column density was fixed density at $N_{\rm H} = 5.26 \times
10^{20} \rm \ cm^{-2}$ (Dickey \& Lockman 1990). 
All model parameters will be referred to in the source frame.  

As an initial assessment of the form of the spectrum (with the
effect of the instrumental response accounted for), in \figplrat we show
the ratio of the data to a model consisting of a simple power-law continuum
with the photon index, $\Gamma$, fixed at the typical value of 1.9
(e.g. Nandra \& Pounds 1994; Dadina 2008), 
modified by a uniform absorber. 
We will give spectral parameters and statistical errors 
below, when a full
fit to the data is discussed.
The spectral ratio shows that there is
an excess of flux at the lowest energies and that there is still
considerable curvature in the spectrum between $\sim 4-8$~keV. We know
from previous observations of NGC~2992
that the soft flux excess is likely to be
due to a combination of Thomson-thin scattering plus optically-thin
thermal emission in the warm extended (kpc-scale) zone that has been directly
imaged by \chandra in NGC~2992 (Colbert \etal 2005) and in other nearby
Seyfert galaxies (e.g. Morse \etal 1995). The scattered power-law continuum
from this region (typically a few percent of the flux of the intrinsic
power-law continuum), when combined with the intrinsic power-law emission
is mathematically indistinguishable from a partial-covering model and 
could potentially account for at least some of the curvature in the 
$\sim 4-8$~keV range.
Otherwise, the residuals in the $\sim4-8$~keV range are similar to
those that one would expect from a relativistically broadened Fe~K line.
The characteristic signature of the narrow core of the Fe~K 
line at $\sim 6.4$~keV
is also evident in the residuals. However, as explained 
in \S\ref{idvsin} and \S\ref{fekresults}, the
distinction between this component and any possible relativistically-broadened
component may be model-dependent. \figplrat also shows the ratio of the
summed MOS1 and MOS2 data to the same model and it can be seen that the MOS data
are consistent with the pn data within the statistical errors.
Since the effects of pile-up are worse for the MOS than the pn,
this indicates that pile-up has been mitigated.

\subsection{Baseline Model}
\label{baselinemodel}

\figcxmorata shows the data to model ratio 
(above 4~keV) when the
absorber in the simple model above was 
replaced with a partial covering model and
the photon index of the power-law
continuum was allowed to float (the fit was still performed
in the 0.5--10~keV band). 
For the data/model residuals shown in \figcxmorata
the data in the 5.0-7.5 keV band, where the Fe~K
line emission would dominate, were
omitted (and  gave  $\chi^2 = 234.6$ for 169 degrees of
freedom). 
However, when the full 0.5--10~keV band data
are taken into account, a poor fit was obtained,
with $\chi^2 = 464.5$ for 233 degrees of
freedom.
Next, we added two unresolved narrow Gaussian components to model the 
distant-matter \feka and \fekbeta line emission.
The intrinsic widths ($\sigma_N$) of the narrow Fe~K lines were tied together and fixed 
at a value of 5 eV, much less than the pn spectral resolution 
($\sim$ 150 eV at 6 keV). The centroid energy of the  \feka line 
($E_{\rm N}$) and its intensity ($I_{\rm N}$) were allowed to
float. The energy of the Fe~$K\beta$ line was 
fixed at 7.058 keV, with the line flux set equal to 13.5\% of the \feka as expected for 
neutral Fe (Kallman \etal 2004). 
A significant improvement in the $\chi^2$ 
was obtained upon adding these two narrow lines 
($\Delta \chi^2 = 107.8$ for two extra parameters). Note that
when an additional broad line is included, the narrow core of
the \fekalfa line becomes statistically less significant (see \S\ref{fekresults}).

\figcxmoratb  shows the residuals in the Fe~K band after fitting 
the narrow Fe K$\alpha$ and K$\beta$ lines. However, it can be seen
that there are still some residuals 
in the Fe K band indicative of additional, broader, line emission
{\it even though some complex absorption is already included}. This excess could 
be associated with the line emission from the inner regions of an accretion disk.  
However, in principle the residuals could alternatively be due to a continuum
that is more complex than has been modeled here, and this type of degeneracy is
still a matter of considerable debate (e.g. see Miller \etal 2009;
Turner \& Miller 2009,
and references therein). 
For example, complex absorption and/or reflection in ionized
matter could mimic relativistic Fe~K line emission for data that has limited 
signal-to-noise and spectral resolution. 
Although Guainazzi \& Bianchi (2007) reported
the detection of four discrete absorption lines 
in the RGS data from the \xmm observation reported here
(C~{\sc vi}~Ly$\beta$, O~{\sc vii}~He$\alpha$(r), O~{\sc viii}~Ly$\alpha$,
and  O~{\sc viii}~Ly$\beta$), the statistical significance of
the lines was low.
We attempted to model the \xmm pn data without a broad \fekalfa relativistic
line component, using a photoionized absorber instead to account
for the apparent wings of the \fekalfa line. However, we found
that such models (which included additional cold absorption, as
well as a Compton reflection continuum), always preferred the
ionization parameter to be driven to low values (rendering
the absorber only mildly unionized), and left sharp 
and statistically significant residuals over
the narrow energy range of $\sim 5.5-6.5$~keV. This is also
evident from \figcxmoratp: the ``excess'' that forms the apparent
red wing of the \fekalfa line is simply too localized in
energy to be explained by photoionized absorption, or for that
matter, a complex continuum in general. 
Indeed, we found that ionized disk reflection models
(e.g. Ross \& Fabian 2005) predict strong soft X-ray emission line
features that are not present in the data, and the amount of 
relativistic blurring required to diminish their amplitudes
makes the \fekalfa line far too broad to account for the data.
The case of NGC~2992 thus contrasts with the
situation for MCG~$-$6-30-15 in which the apparent red wing
of the \fekalfa line is {\it so} broad that it is fairly easy
to mimic with a complex continuum (e.g. Miller \etal 2008, 2009).

We emphasize that the importance of the
\xmm observation discussed here is that it is the {\it only} high-flux state
observation of NGC~2992 with CCD-spectral resolution. All other historical X-ray data 
for NGC~2992 
(see \S\ref{historicaldata})
has inferior spectral resolution in the Fe~K band (the source has
never been observed with the \chandra gratings).
Nevertheless, the signal-to-noise is limited and does not
warrant allowing black-hole spin to be a free parameter
when modeling the broad \fekalfa line so
we used the simple model {\sc diskline} in XSPEC 
for emission from a disk around
a Schwarzschild black hole (see Fabian \etal 1989). 
The parameters of this disk-line model are the rest-frame
line energy ($E_{0}$); the inner and outer radii of the disk  
($R_{\rm in}$ and $R_{\rm out}$, respectively, in units of $R_{g}$); 
the power-law index of the line emissivity ($q$, where the line emissivity is
parameterized as a single power-law function, $r^{q}$); 
the inclination angle of the disk normal with respect to the line-of-sight 
of the observer ($\theta_{\rm obs}$); and the integrated intensity of the line, $I_{\rm disk}$. 
There are still too many parameters to constrain, given the degeneracies, so
we fixed the inner radius at $6R_{g}$
(the marginally stable orbital radius for a Schwarzschild black hole), 
the emissivity at $q = -3$, and $E_{0}$ at 6.4 keV 
(corresponding to neutral Fe). For completeness, we also included 
a Fe K$\beta$ disk line component with no additional free parameters, assuming the
same Fe~$K\beta/K\alpha$ ratio as for the narrow Fe~K line core. 
If Fe is less ionized than Fe~{\sc xvii} the Fe~$K\beta$ line {\it must} 
be produced. 

In the final model we also included an optically-thin thermal emission
component (using the {\sc APEC} model in XSPEC), which we know exists
from direct imaging (see Colbert \etal 2005). The relative contribution of
the thermal component to
the total \xmm spectrum depends on the amplitude of the
nuclear continuum because the
thermal component cannot respond to variability in the nuclear continuum because
of its origin in a zone that is more than 150~pc in size. As expected,
since the nuclear continuum level during the \xmm observation was at a 
historical
high, preliminary spectral fitting showed that the
relative level of the thermal emission was too low to constrain its temperature.
\suzaku has provided the most sensitive measurement of this temperature 
($kT = 0.656^{+0.088}_{-0.061}$~keV -- see Yaqoob \etal 2007)
so in our baseline model we fixed the temperature at the \suzaku value of
0.656~keV. The normalization of the thermal component was allowed to be free.
However, we could only obtain upper limits on
the intrinsic, absorption-corrected,
luminosity of this soft thermal component  of $L_{\rm apec} <0.9 \times
10^{40} \rm \ erg \ s^{-1}$, which is consistent with the
corresponding luminosity measured by \suzakup, namely 
$1.18^{+0.36}_{-0.45} \times 10^{40} \rm \ erg \ s^{-1}$ (Yaqoob \etal 2007). 

In our baseline model
we also included a Compton-reflection 
component using the model {\sc pexrav} in XSPEC
(see  Magdziarz \& Zdziarski 1995).
Although the \xmm data do not have 
a sufficiently wide bandpass
to constrain any possible reflection component well, it can potentially
affect deduced Fe~K line parameters. The
reflection continuum affects only the $6-10$~keV band, 
so the results must be 
interpreted with caution.
The only parameter we allowed to be free in the reflection model 
was the so-called reflection fraction, $R$, which is the
normalization of the reflection continuum relative to the case of a 
steady state,
centrally-illuminated, neutral disk subtending a solid angle of 2$\pi$ at 
the X-ray source. We note that our modeling of a possible reflection
continuum should be interpreted as an empirical parameterization
only because the disk geometry may not be appropriate for all
of the Compton reflection.

In summary, our baseline model consists of 12 free parameters: 
the normalization of the power-law continuum, its slope ($\Gamma$),
the intrinsic absorbing column density
($N_H$), the covering fraction ($f_c$), 
$E_{\rm N}$, $\sigma_N$, $I_{\rm N}$, 
the reflection fraction ($R$), $\theta_{\rm obs}$, $R_{\rm out}$, 
$I_{\rm disk}$, and 
the normalization of the optically-thin thermal continuum component. 
We interpret the partial covering model in terms of a 
fully covered X-ray source plus a fraction ($f_{s}$)
of the intrinsic continuum scattered into the line of sight by a warm,
optically-thin scattering zone, so that $f_{s} = (1-f_{c})/f_{c}$.
In this scenario, if the scattering zone has a
Thomson depth $\tau_{\rm es}$, and
if the portion of the scattering zone that is visible to the observer
subtends a solid angle $\Omega$ at the X-ray source, $f_{s} = \tau_{\rm es} (\Omega/4\pi)$.
This baseline model gives an excellent fit to the \xmm data
($\chi^2= 282.5$ for 225 degrees of freedom).
The best-fitting parameters and their statistical errors are 
shown in \tablespfit and will be discussed in detail in \S\ref{results}.

\figbestmodel shows the best-fitting
baseline model spectrum and the data/model
residuals. The data/model residuals for the same model are also
shown in \figcxmoratc for just the 4--9~keV band in order to
display the Fe~K region more clearly. It can be seen
that there are still some line-like residuals in the $\sim 6.5-7.5$~keV
band. These are likely to be due to Fe~{\sc xxvi}~Ly$\alpha$ and
Ni~$K\alpha$ emission lines which have 
rest-frame energies of 6.966~keV (Pike \etal 1996)
and 7.472~keV (Bearden 1967) respectively.
However, these features are not statistically significant: when modeled
with narrow Gaussian components with centroid energies fixed
at the above values, and with widths fixed at 5~eV, the residuals
vanish but the overall value of $\chi^{2}$ is reduced by only 8.6
for the addition of two free parameters. We obtained 
 EW$=24_{-16}^{+15}$~eV for the \feklya line and
for the Ni~$K\alpha$ line we obtained only an upper limit of 34~eV for the EW.
Thereafter,
we did not include the \feklya and Ni~$K\alpha$ lines in our baseline model.
 
\figrgsoverlay shows the unfolded pn spectrum and best-fitting continuum model.
Overlaid on this are the unfolded RGS1 and RGS2 data (the Fe~K emission-line
components have been removed to prevent the formation of artificially narrow
spectral features in the unfolded pn spectrum). No adjustment was made to
any relative normalizations for overlaying the RGS data and the excellent
agreement between the RGS and pn data shows
that the effects of pile-up in the pn data have been successfully mitigated
in the overlapping energy band.
It can also be seen that the signal-to-noise of the RGS data is
unfortunately very poor.
Since the count rate for the pn spectrum is so much higher
than the count rates for the  RGS spectra, spectral fits
would be overwhelmingly dominated by the pn data. For these reasons we did not
consider the RGS data further.

\subsection{Broad versus Narrow Fe~K Line}
\label{idvsin}

Here we give the results
of investigating the extent to which the broad Fe K disk-line component and
the narrow, distant-matter Fe K line (e.g. see Yaqoob \& Padmanabhan 2004)
may be decoupled from
modeling of the \xmm data.
Using the 
best-fitting baseline model (\S\ref{baselinemodel} and \tablespfitp),
we constructed two-parameter joint confidence contours of the disk-line
intensity ($I_{\rm disk}$) versus the narrow-line intensity ($I_{N}$),
whilst allowing the other 10 free parameters of the model to remain free.
\figidvsin shows the resulting 68\%, 90\%, and 99\% confidence contours.
For comparison, the corresponding confidence contours 
obtained by Yaqoob \etal (2007) from \suzaku data are shown
in \figidvsinp. 
So far, \suzaku has provided the best measurement of the
distant-matter \fekalfa line in NGC~2992 because the continuum
happened to be low and because of the high throughput and 
sensitivity of \suzakup. 
It can be seen that at 99\% confidence the
narrow \feka line intensity 
from the \xmm data could formally be zero, but values of up to
a factor of $\sim 4$ greater than 
the \suzaku value of $\sim 2.5 \times 10^{-5} \
\rm photons \ cm^{-2} \ s^{-1}$ (Yaqoob \etal 2007)
 are not ruled out. In the \xmm
data the narrow \feka line is marginal because
it is weak against the much higher continuum compared with
that in the \suzaku data. On the other hand, the broad \feka line
component is strong in the \xmm data 
and the intensity is required to be non-zero at a
high level of significance in the context of the baseline model
fitted here. Indeed, the intensity of the broad line appears to be
an order of magnitude larger than that measured by \suzaku (Yaqoob \etal 2007)
and \figidvsin shows that the 99\% confidence contours
for the \xmm and \suzaku data
are mutually exclusive, indicating strong variability of the
broad Fe~K line component.

\section{Detailed Results}
\label{results}

We now discuss in detail the results that
were obtained from fitting the
baseline model (\S\ref{baselinemodel}) to the
\xmm pn data for NGC~2992 (see \tablespfitp).

\subsection{Continuum}
\label{contresults}
We obtained $\Gamma = 1.83^{+0.06}_{-0.04}$   for the power-law
photon index of the intrinsic continuum
and a column density of $6.45^{+0.25}_{-0.17} \ \times  10^{21} \rm \ cm^{-2}$ 
for the line-of-sight X-ray obscuration
(see \tablespfitp).
A comparison with historical data is given in \S\ref{historicaldata}.
\tablespfit shows that we measured $(1.3 \pm 0.2)\%$ for the continuum scattered into the
line-of-sight, as a percentage of the  direct line-of-sight continuum
before absorption. \suzaku data provided one of the
best measurements of the scattered continuum (Yaqoob \etal 2007), 
again because of
\suzakup's sensitivity and because of the low level of the nuclear continuum
at the time of the \suzaku observations. 
The corresponding scattered percentage 
from \suzaku was
larger ($7.3 \pm 2.1\%$)
than that measured from the \xmm data 
because
the scattered continuum originates in an extended zone larger than 150~pc
(Colbert \etal 2005) and
the direct continuum during the \xmm observation was at least a 
factor of 8 greater than that during the \suzaku observations. Clearly,
some variability in the absolute
level of the scattered continuum is indicated (a factor of
$\sim 1.5$), but it cannot respond to short-term variations
in the direct continuum and its absolute luminosity must correspond
to an appropriate historically-averaged level of the intrinsic X-ray continuum.
In Yaqoob \etal (2007) we deduced a value of the Thomson depth for
the warm scattering zone of $\tau_{\rm es} \sim 0.1$, and the \xmm data
are consistent with that, given that the intrinsic continuum during
the \xmm observation is likely to be a factor $\sim 5-6$ higher than the
continuum level averaged over more than a year (see Murphy \etal 2007).

\subsection{Fe~K Emission Lines}
\label{fekresults}

In \S\ref{idvsin} and \figidvsin we showed that the \xmm data do not allow
deconvolution of the broad, relativistic Fe~K line component from the narrow
line core. This is because the continuum during the \xmm 
observation was high, but the narrow line originates from a larger
region, at least the size of the BLR, so cannot respond to short-term
continuum changes. Therefore, the narrow-line EW was low and the
line parameters were more
difficult to measure than those from the narrow line
in the 2005 \suzaku
data in which the continuum was a factor of $\sim 8$ lower than
in the \xmm observation. 
The \suzaku measurements of the \fekalfa line core parameters
are the best to date because of the sensitivity of \suzaku and
because the EW of the \fekalfa line core was particularly
large during the \suzaku observations (see Yaqoob \etal 2007).
Thus, both the EW and intensity of the narrow Fe~K line
were poorly constrained by the \xmm data, having 
values $49^{+22}_{-27}$~eV
and $5.7^{+2.6}_{-3.1} \ \times 10^{-5} \ \rm photons \ cm^{-2} \ s^{-1}$
respectively (see \tablespfitp). As indicated by the statistical errors, given the degeneracies,
the narrow Fe~K line was not in fact significantly detected in the \xmm data.
However, the \xmm data are consistent with the narrow Fe~K line being present
with the same absolute intensity as in the \suzaku data 
(see \S\ref{prelimfit} and \figidvsinp).
The upper limit on the width of the narrow Fe~K line 
obtained from the \xmm data (FWHM~$<9 \ 155\rm km \ s^{-1}$), was also 
looser
than the corresponding \suzaku constraint (FWHM~$4850~ \rm km \ s^{-1}$).
See Yaqoob \etal (2007) for a discussion of the associated constraints on the
location of the matter responsible for the Fe~K line core.

On the other hand, the \xmm data provide very strong constraints on a possible
broad, relativistic \fekalfa emission-line component because that component appears
to have increased in absolute flux in tandem with the intrinsic
continuum level, compared to the \suzaku data. From simple modeling
of the \xmm data we 
obtained a broad \fekalfa line flux 
of $I_{\rm disk} = 24.1^{+6.3}_{-6.3}
\ \times 10^{-5} \ \rm photons \ cm^{-2} \ s^{-1}$  (\tablespfitp), 
compared to the
\suzaku value of $1.9^{+0.5}_{-1.0}\ \times 10^{-5} \ \rm photons \ cm^{-2} \ s^{-1}$ (Yaqoob \etal 2007). 
The \suzaku data gave the most sensitive 
and highest spectral-resolution measurements
the broad \fekalfa component when the X-ray continuum was in
a low state. 
Thus, an order of magnitude increase in 
the continuum level appears to have been accompanied by
a similar factor increase in
the flux of the broad \fekalfa line.
Our measurement of the broad \fekalfa line flux is consistent,
within the statistical errors, with Brenneman \& Reynolds (2009)
but they did not account for the severe pile-up in the data (and they also
used a different model and a narrower bandpass).

Next, we investigated the sensitivity of the broad \fekalfa line flux in
the \xmm data to the Compton-reflection continuum.
\tablespfit shows that we obtained only an
upper limit on the
best-fitting value for the Compton-reflection continuum
relative normalization of $R <1.16$, but the
\xmm bandpass can only covers 
a small portion of the Compton-reflection continuum.
For the \xmm data we
constructed joint confidence contours 
of the broad \fekalfa line 
intensity, $I_{\rm disk}$ versus $R$. We found that at 99\% confidence,
the lower limit on $I_{\rm disk}$ 
was at least non-zero over the range $0\leq R \leq3$, and was never less than 
$\rm 10^{-4} \ photons \ cm^{-2} \ s^{-1}$. Thus we conclude that 
the broad \fekalfa line intensity is not sensitive to the fitted normalization
of the Compton reflection continuum.

From our simple model
we found that the outer disk radius ($R_{\rm out}$) for the broad \fekalfa line
emission is sensitive to the radial emissivity parameter, $q$.
For $q=-3$, as adopted here, we could only obtain loose constrains
 of $R_{\rm out}$ of $14.7^{+71.4}_{-4.5}R_{g}$ 
(statistical errors correspond to 90\%
confidence for one parameter, or $\Delta \chi^{2} =2.706$).
We note, however, that this solution for $R_{\rm out}$ is a local
minimum only, and the value of $\Delta \chi^{2}$ drops below
2.7 more than once outside the above range. At $R_{\rm out}=6.1R_{g}$
$\Delta \chi^{2}=2.82$, and at $R_{\rm out}=1000R_{g}$,
$\Delta \chi^{2}=1.89$. In other words, $R_{\rm out}$ is very
poorly constrained.
We found that
a steeper radial emissivity law tends to favor larger upper limits
on $R_{\rm out}$ and a flatter emissivity law gives smaller
preferred values on the upper bound on the 
outer radius. For example, with $q=-2$, 
$R_{\rm out}=14.0^{+7.4}_{-3.7}R_{g}$ 
(again the errors correspond to 90\% confidence for one parameter,
but the minimum in $\chi^{2}$ is more robust, giving a lower
limit on $R_{\rm out}$ as well as an upper limit).
We found fairly tight constraints on the disk inclination angle,
 obtaining $\theta_{\rm obs} = 38.0^{+2.7}_{-6.3}$ degrees (\tablespfitp).
This is consistent with the lower limit obtained from the
broad \fekalfa line component in the \suzaku data ($\theta_{\rm obs} >31$ degrees--see Yaqoob \etal 2007). 

\section{Comparison with other historical data}
\label{historicaldata}

In \S\ref{historicalflux} below we discuss
the \xmm observation of NGC~2992 in the general context of the historical
broadband spectra, flux, and variability.
In \S\ref{historicalspectra} we discuss
how the detailed X-ray spectroscopy of the Fe~K band 
in NGC~2992 based on the \xmm data compares
with historical data.

\subsection{X-ray Variability history of NGC~2992}
\label{historicalflux}

Murphy \etal (2007) showed a historical lightcurve of NGC~2992 
spanning $\sim 30$~years, going back to {\it HEAO}-1.
The 2--10~keV flux varied by a factor of over $20$, with 
an \asca observation in 1994 (Weaver \etal 1996) still holding 
the record for 
the lowest 2--10~keV observed ($\sim 4 \times 10^{-12}$ \ergsp).
The \asca data revealed a very prominent, narrow \feka emission
line, with an equivalent width (EW) $\sim$ 500 eV.
The narrow-line intensity had not responded to the declining continuum,
indicating that this particular
line component likely arose in matter far from the
supermassive black hole (possibly in the putative obscuring torus).
Later, two \bepposax observations in the period 1997--1998 (Gilli \etal 2000)
yielded data corresponding
to continuum levels that were different by an order of magnitude to
each other, revealing Fe~K line emission that was more complicated
than previous data had indicated. There was evidence for
possible contributions from
both the accretion disk and distant matter, as well as 
line components from highly ionized Fe.
More recently,  a one-year monitoring campaign of NGC~2992
with {\it RXTE} beginning in 2005
revealed that strong continuum flux variations by a factor of
up to $\sim 11$ (from $\sim 0.8-9 \times 10^{-11} \
\rm erg \ cm^{-2} \ s^{-1}$ in the 2--10~keV band),
actually occur frequently, on timescales
of days to weeks (Murphy \etal 2007).
Additionally, during high-flux periods the spectrum was dominated by a
highly redshifted and
broadened \feka emission line.
Murphy \etal (2007) interpreted this redshifted and
broadened \feka line in the high continuum state in terms of the
temporary enhancement
of line emission from the inner region of
the accretion disk ($< 100 R_{g}$),
where strong gravitational and Doppler effects are important.
Still, there were considerable degeneracies in the \bepposax and
\rxte data.
Meanwhile, \suzaku  observations of NGC 2992 at the end of 2005
caught the source in relatively low state, with a 2 -- 10 keV flux of
1.1 $\times 10^{-11}$ \ergs (Yaqoob \etal 2007).
In particular the complex \feka line emission profile,
consisting of a low-level, ``persistent'' accretion-disk
component and a distant-matter component, was observed and could be decoupled
at a confidence level greater than $3\sigma$.
Hard X-ray data taken by {\it INTEGRAL} (in 2005 May)
and {\it Swift} (in 2005--2006) showed that the
X-ray continuum extends to at least $\sim 200$~keV,
but the data were not sensitive enough to perform further
detailed spectroscopy of the Fe~K line complex (Beckmann \etal 2007).
The \xmm observation reported in the present paper 
represents the highest 2--10~keV flux state of NGC~2992 ever observed.
 
\subsection{Historical X-ray Spectral results}
\label{historicalspectra}

We now investigate whether historical X-ray spectra for NGC~2992 
can be understood in terms of the baseline model that has been
used to fit the \xmm data. Only \suzaku data (Yaqoob \etal 2007)
have spectral resolution and signal-to-noise that are comparable
to the \xmm data, and comparisons between the two data sets have
been given throughout the present paper. The two data sets are
consistent with both narrow and broad \fekalfa lines being present,
but in the \xmm data (high-flux state) the broad line dominated,
and in the \suzaku data (low-flux state), the narrow line dominated.
Unfortunately, the spectral resolution and/or signal-to-noise of most
of the other historical data is not sufficient to perform
meaningful fits with a dual (broad plus narrow) \fekalfa line model.
Thus, we performed such an analysis only on the historical 1994 \asca spectrum
and the 1997 and 1998 \bepposax spectra. It is difficult
to compare our fits to the \xmm data with published results
in the literature for the \asca and \bepposax data because 
different models have been used to fit the data.
We therefore used the same baseline model as that used to fit the \xmm data
(\S\ref{baselinemodel}) but fixed several parameters due to the poorer
signal-to-noise (and in the case of the \bepposax data poorer
spectral resolution as well). The inclination angle 
($\theta_{\rm obs}$) of the accretion
disk, and the outer radius of emission on the disk ($R_{\rm out}$)
were fixed at the respective best-fitting values from the fit
to the \xmm data (\tablehistfitp). 
The \asca data were fitted in the range 0.5--10~keV and the
\bepposax data were fitted in the 1--100~keV range using both
the MECS and PDS instruments (see Gilli \etal 2000).
Neither the \asca nor the \bepposax data could constrain the
luminosity of the soft thermal emission component (\S\ref{baselinemodel}),
so it was not included in the fits, but it was possible to
obtain 
an upper limit on its luminosity for the \asca data.
All of 
the spectral-fitting results for the \asca and \bepposax
data are shown in \tablehistfitp.

Within the statistical errors, the power-law slope appears
to be consistent amongst the different data sets, including the
\xmm data. However, variability in the slope cannot be ruled
out. On the other hand, the column density measured from the
\xmm data is formally inconsistent with that measured from
the two \bepposax observations. However, the \bepposax
data only extend down to 1~keV and the spectral resolution is
poorer than than the \xmm data so any variability in the column
density must be interpreted with caution.

As might be expected, the broad \fekalfa line intensity could not
be constrained by the \asca data and the low-state \bepposax data,
and only upper limits could be obtained. However, for the
high-state (1998) \bepposax data we measured the broad-line intensity
to be $14.7^{+9.1}_{-9.8} \times 10^{-5} \rm \ photons \ cm^{-2} \
s^{-1}$. This is lower than the 
corresponding intensity measured during the \xmm observation
and higher than the that measured the \suzaku observation.
However, the 99\% confidence, two-parameter broad-line
intensity contours from both \xmm and \suzaku overlap
the \bepposax measurement (see \figidvsinp).  
Nevertheless, the collective data are not inconsistent with
the broad-line flux responding to variability in the continuum flux
(see \tablespfit and \tablehistfit for continuum fluxes).

We confirmed the large EW of the narrow \fekalfa line at $\sim 6.4$~keV
during the \asca observation (see \tablehistfitp),
when the X-ray spectrum was dominated by a scattered
continuum and the narrow \fekalfa line, neither of which had responded
to the dramatic decline in the continuum flux compared to
earlier epochs (see Murphy \etal 2007).
However, the behavior of the narrow \fekalfa line 
during the \bepposax observations is puzzling. In the low-state 1997
\bepposax observation the narrow line at 6.4~keV was not detected
(but the upper limit on its flux,
shown in \tablehistfitp, is consistent with historical data).
However, there is instead a statistically significant
($\Delta\chi^{2}=30.2$ for the addition of two free parameters) narrow
emission line at $6.60^{+0.12}_{-0.06}$~keV (see \tablehistfit for parameters).
During the high-state (1998) \bepposax observation the narrow line
at 6.4~keV was again marginally detected ($\Delta\chi^{2}=1.7$),
but its flux is formally consistent with that during the \asca
observation (for 90\% confidence, one parameter). Moreover,
in the 1998 \bepposax data there are residuals that can be
modeled by
an additional ionized Fe emission line at $7.08^{+0.31}_{-0.39}$~keV,
but the statistical significance is low ($\Delta\chi^{2}=3.5$).
All of these unusual
features have already been noted in earlier analyses of the 
1997 and 1998 \bepposax data (Gilli \etal 2000). 

\subsection{Variation of the Reflection Spectrum}
\label{kyfitting}

{We note that an increase in the flux of the \fekalfa line,
as observed during the \xmm observation, compared to historical data,
should be accompanied by a corresponding increase in the magnitude
of the associated continuum reflection. This is because the 
relativistic 
component of the \fekalfa line and the reflection continuum
both originate
from reprocessing the same intrinsic continuum
at the same physical locations so the 
flux of the line relative to the reflection continuum 
should not itself
vary as the intrinsic continuum varies.
We have applied the reflection model {\sc kyl1cr}, which is
publicly available for the {\tt XSPEC} package, to the \xmm and \suzaku 
data. The model calculates the reflection continuum and the associated 
\fekalfa and \fekbeta emission lines self-consistently for a neutral disk,
including relativistic blurring applied to continuum and emission lines.
Full details of the model and its parameters can be found in 
Dov\v{c}iak \etal (2004b). The model is only valid above 2~keV so we
fitted both data sets in the 2--10~keV band. The purpose here is to
simply assess whether these two data sets, representing low (\suzakup) and
high (\xmmp) continuum states are at least consistent with the entire reflection spectrum (continuum plus broad emission lines) following the order-of-magnitude change in continuum flux between the \suzaku and \xmm observations. 
Since the fits were performed only above 2~keV we fixed the absorbing column density, $N_{H}$, and the scattering fraction, $f_{s}$, at the best-fitting values 
for \xmmp (\tablespfitp) and \suzaku (Yaqoob \etal 2007). 
The soft thermal emission component was omitted as it has no impact in the 
fitted energy band. The inner and outer radii of the disk were 
fixed at $6R_{g}$ and $1000R_{g}$, and the disk
radial emissivity index, $q$, was fixed at $-3$. The black-hole spin was 
fixed at a value of zero. 
The {\sc kyl1cr} model includes the \fekbeta as well as the \fekalfa 
emission line.
Narrow gaussian emission lines to model the 
distant-matter \fekalfa and \fekbeta emission
lines were still included but the 
centroid energies and intrinsic line widths 
were fixed at the best-fitting
values from the appropriate empirical 
fits in \tablespfit and Yaqoob \etal (2007). Our conclusions
with respect to variability of the
disk-reflection features are not sensitive to these parameters.
This left a total five free parameters for the error analysis, namely,
the intrinsic power-law normalization and photon index ($\Gamma$), 
the reflection spectrum normalization,
the disk inclination angle, and the flux of the narrow \fekalfa line
from distant-matter. Note that the definition of the 
normalization of the reflection spectrum here is peculiar to the {\sc kyl1cr}
model, being the flux of the reflection spectrum at 3~keV in units of
${\rm photons \ cm^{-2} \ s^{-1} \ keV^{-1}}$ 
(see Dov\v{c}iak \etal 2004b for details). }

{The results of the {\sc kyl1cr} fits
are shown in \tablekymodel and the best-fitting models for the \xmm and \suzaku data are shown in \figkya and \figkyb respectively. 
The 4--10~keV band is shown in the plots in order to exhibit detail in
the complex Fe~K region of the spectra.
The $\chi^{2}$ and reduced $\chi^{2}$ values in \tablekymodel
show that the fits are essentially as good as the previous spectral
fits with {\sc pexrav} reported 
for the \xmm data in \tablespfit and for the \suzaku data
in Yaqoob \etal (2007).
A direct comparison of the reduced $\chi^{2}$ values 
is given in the
footnotes to \tablekymodelp.
In \figky
the black spectra show the summed direct and reflected models and the 
red spectra show the reflection models only (continuum plus lines). It can 
seen from both the numerical results in \tablekymodel and the 
spectra in \figky that indeed, the \xmm and \suzaku data are 
consistent with the reflection spectra (continuum plus lines) 
following the intrinsic continuum variability. Both data sets are 
consistent with the same inclination angle, $\sim 40^{\circ}$.
Further observation and more detailed self-consistent modeling
of the emission-line and continuum components is essential
and some of the authors of the present
paper have obtained observing time on NGC~2992
for an extended
monitoring campaign with \xmmp, along with a quasi-simultaneous
\chandra high-energy transmission grating observation.
The new data promise to provide new insight into the 
properties and origins of all of the \fekalfa emission-line 
and continuum components, and into the structures responsible for
producing them.}

\section{Discussion and Conclusions}
\label{conclusions}

In the past three decades, the Seyfert galaxy 
NGC~2992 has shown X-ray continuum flux variations 
of more than a factor of 20. 
We have reported the results of an \xmm observation of NGC~2992
performed in 2003,
during which the source was found to have the highest 2--10~keV
flux ($9.4 \times 10^{-11} \ \rm erg \ cm^{-2} \ s^{-1}$)
compared to previous historical values. No previous X-ray
astronomy mission to date has observed NGC~2992 in a high state
(2--10~keV flux greater than $\sim 7 \times 10^{-11} \ \rm 
erg \ cm^{-2} \ s^{-1}$) with CCD spectral resolution 
($\sim 7000 \ \rm km \ s^{-1}$ FWHM at 6.4~keV).
The best spectral resolution available for any previous 
high-state data set for NGC~2992 was a factor of $\sim 3$ worse than that
for the \xmm data in the
Fe~K band. The rather unique \xmm pn 
spectrum of NGC~2992
has a broad feature in the $\sim 5-7$~keV band that can be
interpreted as relativistic \fekalfa line emission. 
Its flux is an order of magnitude larger
than that found in \suzaku data (obtained in 2005), when the 
2--10~keV continuum flux of NGC~2992 was a factor of $\sim 8$ less than that
during the \xmm observation. Although the detailed \fekalfa 
line parameters obtained from the \xmm data are model-dependent,
it appears that the absolute 
luminosity of the broad \fekalfa line and the
associated reflection continuum increase as the
continuum luminosity increases. 

The observation of {\it variable} broad, relativistic \fekalfa line emission
in AGN is rare, and the observation of such variability in response
to X-ray continuum variability is even rarer.
In the light-bending
model (Fabian \& Vaughan 2003; Miniutti \& Fabian 2004) invoked to account for
the {\it non-variability} of the broad \fekalfa line in MCG~$-$6-30-15,
the X-ray continuum variability for an observer at infinity
is due entirely to relativistic effects as the X-ray source
physically changes position relative to the black hole. The
disk, being much closer to the X-ray source and black hole
than the distant observer, is
then not subject to large variability in illumination by the
X-ray continuum and therefore produces an \fekalfa emission line
that is not variable. In this model there can of course 
be no intrinsic variability of the X-ray source, otherwise
the \fekalfa line would be variable. The \xmm data for NGC~2992,
when considered in the context of historical data,
therefore suggest that the light-bending scenario is not relevant
for this AGN, implying that the large-amplitude X-ray continuum 
variability is {\it intrinsic} to the X-ray source. Further,
if there were X-ray continuum variability that had {\it both}
an intrinsic origin, and one due to relative motion of the
X-ray source and black hole, the broad \fekalfa line and continuum
variability would not be related in a simple way.

The high level of the intrinsic continuum during the \xmm observation
reported here swamped the features that originate in circumnuclear
matter that is more extended than the X-ray source because these features
did not respond to the large-amplitude change in the intrinsic
continuum that must have occurred prior to the start of the 
\xmm observation. These features 
(namely the optically-thin thermal emission, the scattered
intrinsic continuum, and the narrow core of the \fekalfa line) have
been better studied in historical spectra taken during low-continuum states
(e.g. Weaver \etal 1996; Gilli \etal 2000; Yaqoob \etal 2007).
  
We thank the referee for his/her useful comments. 
The authors thank the \xmm instrument teams and operations staff 
for making the observation of NGC~2992.
This research made use of the HEASARC online data archive services, supported
by NASA/GSFC.
The work was supported by Chinese NSF through Grant 10773010/10825312, and the 
Knowledge Innovation Program of CAS (Grant No. KJCX2-YW-T05).

\newpage
\begin{table}[!htb]
\centerline{Table 1. Spectral fitting results for NGC~2992 {\it XMM-Newton} data.} 
\begin{center}
\begin{tabular}{lr}
\\
\hline
\\
$\chi^{2}$ / degrees of freedom & 282.5/225 \\
& \\
$L_{\rm APEC}$ (10$^{40}$ erg s$^{-1}$) & $<0.9$ \\
$\Gamma$ & $1.83^{+0.06}_{-0.04}$ \\
$N_{H} \rm \ (10^{21} \ cm^{-2})$ & $6.45^{+0.25}_{-0.17}$ \\
$f_{\rm s}$ (scattered fraction) & $0.013^{+0.002}_{-0.002}$ \\
$\theta_{\rm obs}$ (degrees) & $38.0^{+2.7}_{-6.3}$ \\
Reflection fraction, R & $0.40 (<1.16)$ \\
Outer Radius of disk, R$_{\rm out}$ & $14.7^{+71.4}_{-4.5}$ \\
$I_{\rm disk} \rm \ [Fe~K\alpha]$ ($\rm 10^{-5} \ photons \ cm^{-2} \ s^{-1}$)
& $24.1^{+6.3}_{-6.3}$ \\
$\rm EW_{disk} \ [Fe~K\alpha]$ (eV) & $255^{+67}_{-67}$ \\
$E_{N} \rm \ [Fe~K\alpha]$ (keV) & $6.403^{+0.025}_{-0.027}$ \\
$\sigma_{N}$ (keV)  &$<0.083$ \\
FWHM ($\rm km \ s^{-1}$)  & $<9 \ 155$ \\
$I_{\rm N} \rm \ [Fe~K\alpha]$ ($\rm 10^{-5} \ photons \ cm^{-2} \ s^{-1}$)  &
$5.7^{+2.6}_{-3.1}$  \\
$\rm EW_{N} \ [Fe~K\alpha]$ (eV)  & $49^{+22}_{-27}$ \\
$F_{\rm 0.5-2 \ keV}$ ($10^{-11} \rm \ erg \ cm^{-2} \ s^{-1}$)$^{a}$ & 1.8 \\
$F_{\rm 2-10 \ keV}$ ($10^{-11} \rm \ erg \ cm^{-2} \ s^{-1}$)$^{a}$  & 9.5 \\
$L_{\rm 0.5-2 \ keV}$ ($10^{43} \rm \ erg \ s^{-1}$)$^{b}$ & 0.84 \\
$L_{\rm 2-10 \ keV}$ ($10^{43} \rm \ erg \ s^{-1}$)$^{b}$  & 1.3 \\
\\
\hline
\end{tabular}
\end{center}
Statistical errors and upper limits correspond to 90\% confidence
for one interesting parameter
($\Delta \chi^{2} = 2.706$) and were derived with twelve parameters
free. See \S\ref{baselinemodel} for details of the model.
Note that the constraints on $R_{\rm out}$ correspond to
a local minimum only and should be interpreted with
caution (see text for details).
All parameters (except continuum fluxes) refer to the rest frame of NGC~2992.
$^{a}$ Observed-frame fluxes, {\it not} corrected 
for Galactic and intrinsic absorption.
$^{b}$ Intrinsic, rest-frame luminosities, corrected for all
absorption components. 
\end{table}
 
\newpage

\begin{table}[!htb]
\centerline{Spectral fitting results for historical NGC~2992 data.} 
\begin{center}
\begin{tabular}{lrrr}
\hline
\\
Parameter & \asca & \bsax & \bsax \\
& (1994) & (1997--SAX1) & (1998--SAX2) \\
\\
\hline
\\
$\chi^{2}$ / degrees of freedom & 99.2/136 & 189.9/204 & 253.9/248  \\
& \\
$L_{\rm APEC}$ (10$^{40}$ erg s$^{-1}$) & $<3.1$ & \ldots & \ldots \\
$\Gamma$ & $1.50^{+0.72}_{-0.32}$ & $2.08^{+0.34}_{-0.31}$
& $1.79^{+0.07}_{-0.06}$  \\
$N_{H} \rm \ (10^{21} \ cm^{-2})$ &  $9.4^{+5.3}_{-7.1}$  
& $33.5^{+15.8}_{-17.1}$  & $12.4^{+8.0}_{-3.8}$  \\

$f_{\rm s}$ (scattered fraction) & $1.0^{+3.9}_{-0.7}$ 
& $0.35^{+0.32}_{-0.23}$ & $0.18^{+0.38}_{-0.18}$  \\

Reflection fraction, R & \ldots & $2.6^{+4.8}_{-2.1}$
& $0.1^{+0.2}_{-0.1}$  \\


$I_{\rm disk} \rm \ [Fe~K\alpha]$ ($\rm 10^{-5} \ photons \ cm^{-2} \ s^{-1}$)
& $<5.6$ & $<1.8$  & $14.7^{+9.1}_{-9.8}$  \\
$\rm EW_{disk} \ [Fe~K\alpha]$ (eV) & $<1190$  &  $<150$  & $188^{+115}_{-125}$  \\
$E_{N} \rm \ [Fe~K\alpha]$ (keV) & $6.41^{+0.05}_{-0.05}$ & 6.4 (f) & 
6.4 (f) \\
$I_{\rm N} \rm \ [Fe~K\alpha]$ ($\rm 10^{-5} \ photons \ cm^{-2} \ s^{-1}$)  &
$2.6^{+1.4}_{-1.3}$ & $<2.1$ & $2.8^{+3.6}_{-2.8}$  \\
$\rm EW_{N} \ [Fe~K\alpha]$ (eV)  & $506^{+272}_{-253}$ & $<184$ & 
$32^{+39}_{-32}$  \\
Ionized line, $E$ & \ldots & $6.60^{+0.12}_{-0.06}$ & $7.08^{+0.31}_{-0.39}$  \\
Ionized line, $I$ & \ldots & $3.4^{+0.7}_{-0.8}$ & $2.5^{+2.3}_{-2.1}$ \\
Ionized line, $EW$ & \ldots & $494^{+102}_{-116}$ & $35^{+32}_{-29}$  \\
$F_{\rm 0.5-2 \ keV}$ ($10^{-11} \rm \ erg \ cm^{-2} \ s^{-1}$)$^{a}$ & 0.086 
& \ldots & \ldots \\
$F_{\rm 2-10 \ keV}$ ($10^{-11} \rm \ erg \ cm^{-2} \ s^{-1}$)$^{a}$  & 0.44 
& 0.59 & 7.4 \\
$L_{\rm 0.5-2 \ keV}$ ($10^{43} \rm \ erg \ s^{-1}$)$^{b}$ & 0.022 
& \ldots & \ldots \\
$L_{\rm 2-10 \ keV}$ ($10^{43} \rm \ erg \ s^{-1}$)$^{b}$  & 0.060 & 0.092 
& 1.1 \\
\\
\hline
\end{tabular}
\end{center}
Note that results of fits to historical \suzaku data using a
similar model can be found in Yaqoob \etal (2007).
Statistical errors and upper limits correspond to 90\% confidence
for one interesting parameter
($\Delta \chi^{2} = 2.706$) and were derived with eight parameters
free. 
All parameters (except continuum fluxes) refer to the rest frame of NGC~2992.
Some of the disk and distant-matter \fekalfa 
emission-line parameters were frozen --
see \S\ref{historicaldata} for details of the model.
Frozen parameters are indicated by ``(f)''.
$^{a}$ Observed-frame fluxes, {\it not} corrected 
for Galactic and intrinsic absorption.
$^{b}$ Intrinsic, rest-frame luminosities, corrected for all
absorption components. 
\end{table}

\newpage

\newpage
\begin{table}[!htb]
\centerline{Table 3. Spectral fitting results for the {\sc kyl1cr} model.} 
\begin{center}
\begin{tabular}{lrr}
\\
\hline
\\
& \xmm & \suzaku \\
$\chi^{2}$ / degrees of freedom & 245.8/193 & 240.3/267\\
Reduced $\chi^{2}$ $^{a}$ & 1.27 & 0.90 \\
$\Gamma$ & $1.99^{+0.13}_{-0.05}$ & $1.75^{+0.10}_{-0.10}$ \\
$\theta_{\rm obs}$ (degrees) & $40.7^{+3.3}_{-4.6}$ & 
	$39.8^{+4.1}_{-4.7}$ \\
{\sc kyl1cr} normalization $^{b}$ & $1.18^{+0.15}_{-0.16} \times 10^{-4}$ 
	& $1.09^{+0.36}_{-0.48} \times 10^{-5}$ \\
$I_{\rm N} \rm \ [Fe~K\alpha]$ ($\rm 10^{-5} \ photons \ cm^{-2} \ s^{-1}$)  &
$6.1^{+1.6}_{-1.5}$  & $3.1^{+0.3}_{-0.3}$ \\
$\rm EW_{N} \ [Fe~K\alpha]$ (eV)  & $53^{+14}_{-13}$ & $218^{+21}_{-21}$ \\
$F_{\rm 2-10 \ keV}$ ($10^{-11} \rm \ erg \ cm^{-2} \ s^{-1}$)$^{c}$  & 9.5 & 1.1 \\ 
$L_{\rm 2-10 \ keV}$ ($10^{43} \rm \ erg \ s^{-1}$)$^{d}$  &  1.3 & 0.16 \\ 

\\
\hline
\end{tabular}
\end{center}
{
Results of spectral fitting with the relativistic disk model {\sc kyl1cr}
(see \S\ref{kyfitting} for details). 
Statistical errors and upper limits correspond to 90\% confidence
for one interesting parameter
($\Delta \chi^{2} = 2.706$) and were derived with five parameters
free. 
All parameters (except continuum fluxes) refer to the rest frame of NGC~2992.
$^{a}$ For comparison, 
the reduced $\chi^{2}$ values for the fits with {\sc pexrav} were
1.28 and 0.89 for the \xmm (\tablespfitp) and \suzaku 
(Yaqoob \etal 2007) data respectively.
$^{b}$ The definition of the normalization of the reflection spectrum is peculiar to the {\sc kyl1cr} model and is defined as the monochromatic flux at 3~keV in units of ${\rm photons \ cm^{-2} \ s^{-1} \ keV^{-1}}$ (see  Dov\v{c}iak \etal 2004b).
$^{c}$ Observed-frame fluxes, {\it not} corrected 
for Galactic and intrinsic absorption.
$^{d}$ Intrinsic, rest-frame luminosities, corrected for all
absorption components. }
\end{table}
 
\newpage

\begin{figure}[!htb]
 \centerline{
\epsscale{0.5}
     \scalebox{0.6}{\includegraphics[angle=270]{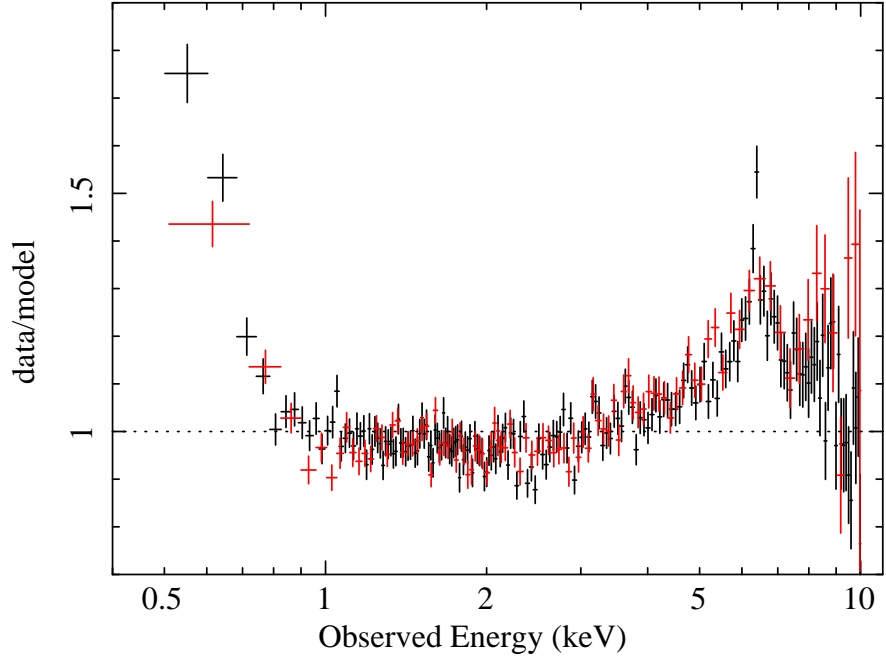}}}
 \caption{
The ratios of the NGC~2992 \xmm EPIC spectra to a model consisting of a simple,
absorbed power-law continuum (see \S\ref{prelimfit}). Black and
red data points correspond to the pn and summed MOS1 and MOS2 data
respectively.
}
\end{figure}

\newpage

\begin{figure}[htb]
 \centerline{
\epsscale{0.5}
\scalebox{0.6}{\includegraphics[angle=270]{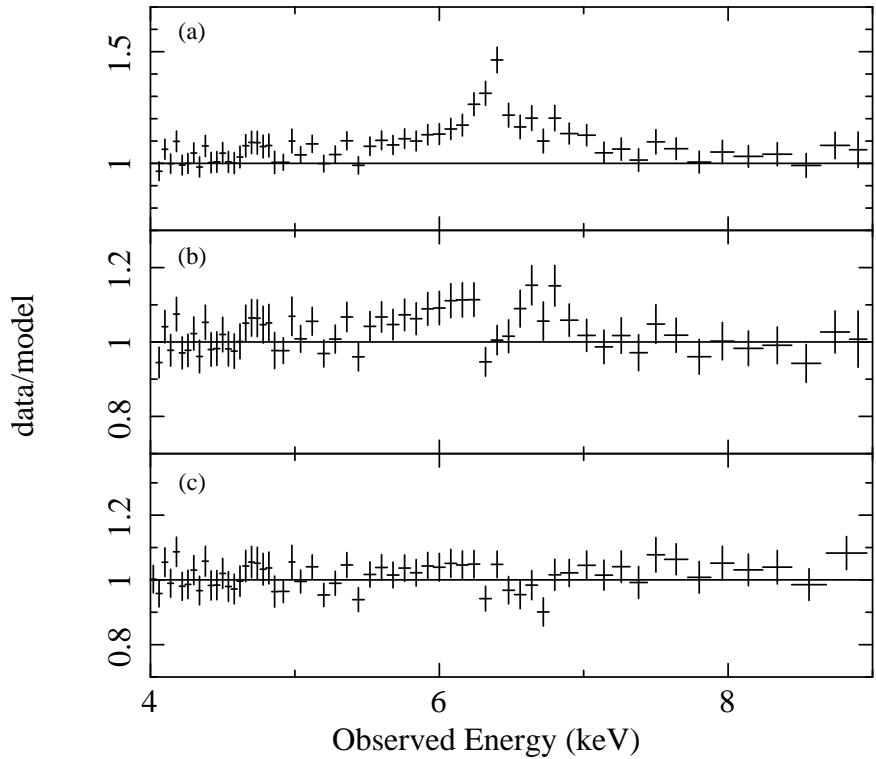}}}
\caption{Data/model ratios for the 
NGC~2992 \xmm pn data in the 4--9~keV band
(see \S\ref{prelimfit} and \S\ref{baselinemodel} for details).
 (a) Data/model ratio after fitting an absorbed power-law
continuum with partial covering.
 (b) Residuals after the narrow K$\alpha$ and K$\beta$ Fe lines 
have been fitted.
 An excess between 5.8-7 keV is still present. 
(c) Residuals after the addition of a broad \feka line, using the 
full baseline model described in \S\ref{baselinemodel} (see also
\tablespfitp).}
\end{figure}

\begin{figure}[htb]
 \centerline{
\epsscale{0.5}
\scalebox{0.6}{\includegraphics[angle=0]{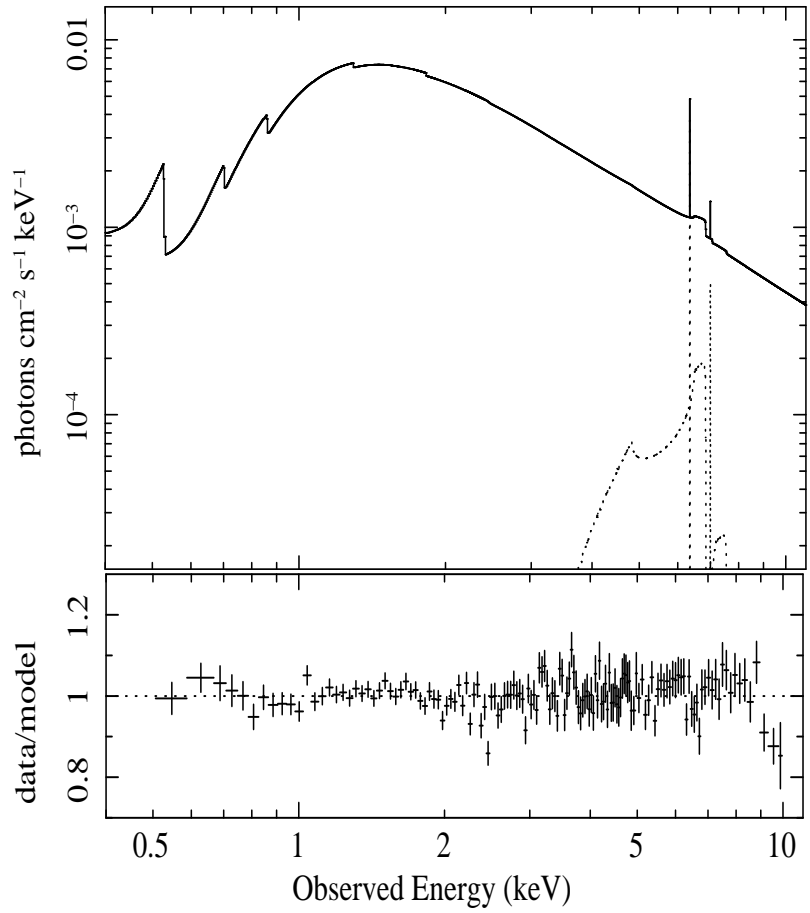}}}
\caption{
The best-fitting baseline model 
to the NGC~2992 \xmm pn data (top panel), 
and the corresponding data/model ratio
(see \S\ref{baselinemodel} and \tablespfit for details). }
\end{figure}

\newpage

\begin{figure}[htb]
 \centerline{
\epsscale{0.5}
\scalebox{0.6}{\includegraphics[angle=270]{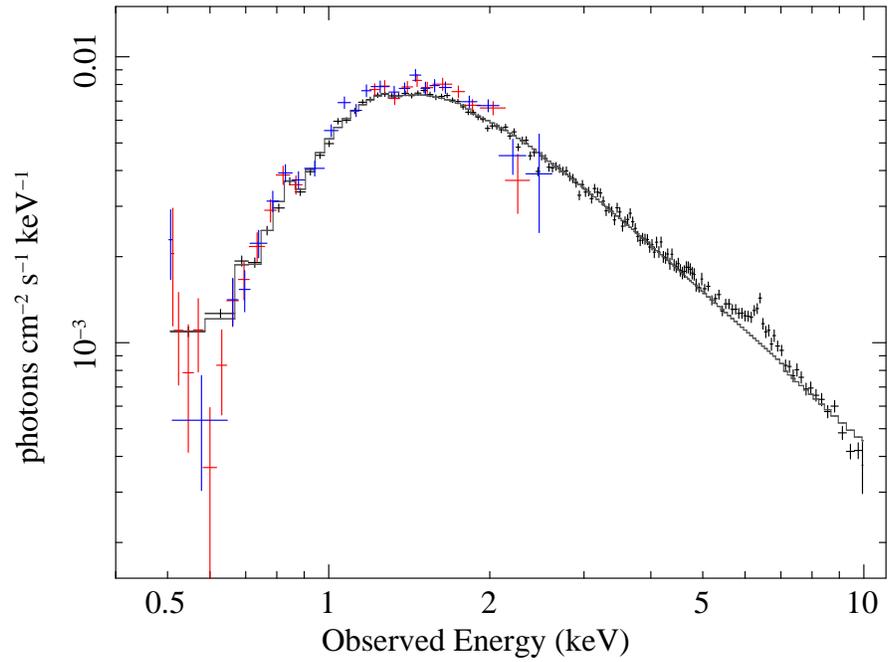}}}
\caption{
The unfolded \xmm pn data for NGC~2992 (using the best-fitting baseline
model described in \S\ref{baselinemodel}, with the emission-line
features removed in order to avoid artificial narrow features
in the unfolded spectrum). Overlaid are the RGS1 (red) and
RGS2 (blue) data, with no relative normalization adjustments.
This shows that effects of pile-up in the pn spectrum have
been successfully mitigated in the overlapping bandpass}. 
\end{figure}

\newpage

\begin{figure}[htb]
 \centerline{
\epsscale{0.6}
\scalebox{0.7}{\includegraphics[angle=0]{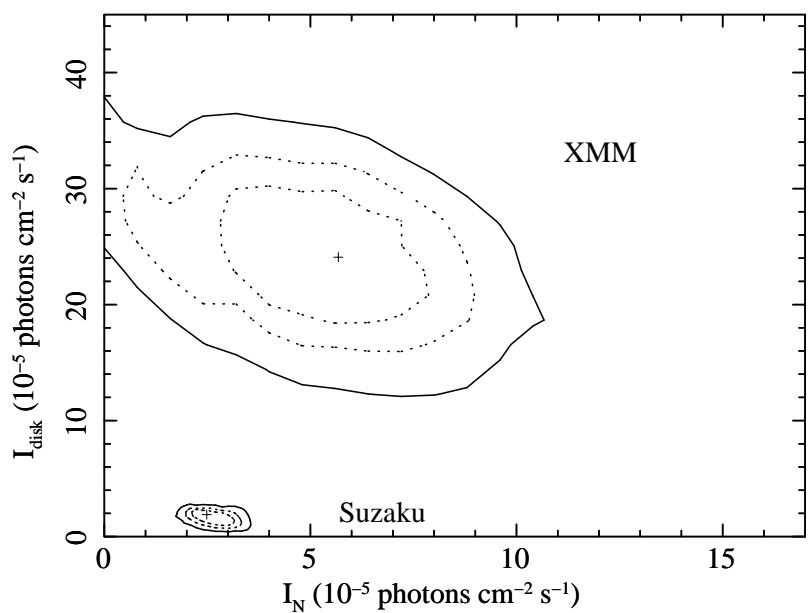}}}
\caption{The 68\%, 90\%, \& 99\% confidence contours of the broad \feka 
line intensity, ($I_{\rm disk}$), versus the intensity of the narrow \feka 
line, ($I_{\rm N}$),
for the NGC~2992 \xmm data (see \S\ref{prelimfit} for details).
Also shown for comparison are the corresponding confidence contours
for the
\suzaku NGC~2992 data reported in Yaqoob \etal (2007).}
\end{figure}

\newpage

\begin{figure}[htb]
 \centerline{
\epsscale{0.6}
\scalebox{0.7}{\includegraphics[angle=0]{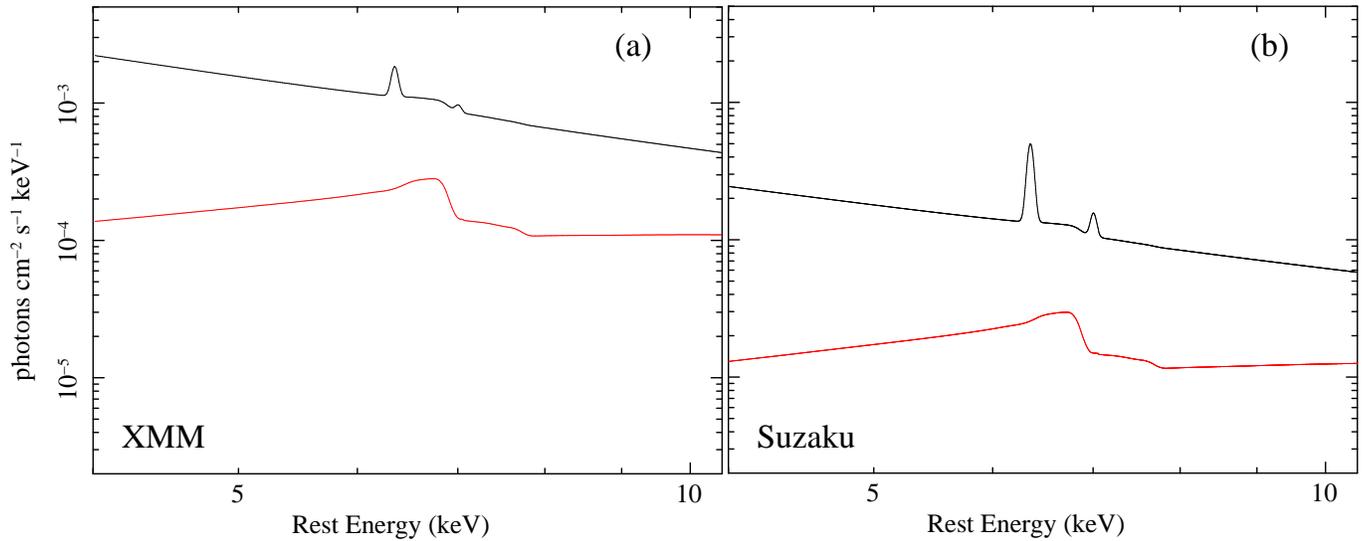}}}
\caption{{Best-fitting {\sc kyl1cr} models to (a) the \xmm data, and
(b) the \suzaku data. The black spectra correspond to the sum of the
 intrinsic power-law continuum plus the reflection continuum and
broad and narrow \fekalfa and \fekbeta emission lines, 
and the red spectra correspond to the reflection components only
(i.e. reflected continuum disk emission lines after relativistic
blurring). The 4--10~keV band is shown in order to exhibit greater detail in the complex Fe~K region of the spectra.}
}
\end{figure}


\newpage

\begin{thebibliography}{}

\bibitem[Arnaud 1996]{arnaud1996}
Arnaud, K. A. 1996, in Astronomical Data Analysis
Software and Systems V, ed. Jacoby, G., \& Barnes, J.
(Astronomical Society of the Pacific), Conference Series, Vol. 101, p. 17 

\bibitem[Ballet 1999]{ballet1999}
Ballet, J. 1999, A\&AS, 135, 371

\bibitem[Bearden 1967]{bearden1967}
Bearden, J. A. 1967, Rev. Mod. Phys., 39, 78

\bibitem[Beckmann, Gehrels, \& Tueller]{beckmann2007}
Beckmann, V., Gehrels, N., \& Tueller, J. 2007, ApJ, 666, 122

\bibitem[Beckwith 2004]{beckwith2004} 
Beckwith, K. \& Done, C. 2004, MNRAS, 352, 353


\bibitem[Brenneman2006]{brenneman2006} 
Brenneman, L. W. \& Reynolds, C. S. 2006, ApJ, 652, 1028

\bibitem[Brenneman2006]{brenneman2009}
Brenneman, L. W. \& Reynolds, C. S. 2009, ApJ, in press (astro-ph/0907.3850)

\bibitem[Colbert \etal 2005]{colbert2005}
Colbert, E. J. M., Strickland, D. K., Veilleux, S., Weaver, K. A.
2005, ApJ, 628, 113 

\bibitem[Dadina 2008]{dadina2008}
Dadina, M. 2008, A\&A, 485, 417

\bibitem[Dickey \& Lockman 1990]{dick1990}
Dickey, J. M., \& Lockman, F. J. 1990, ARA\&A, 28, 215

\bibitem[dovciak2004b]{dovciak2004b}
Dov\v{c}iak, M., \& Karas, V., Martocchia, A., Matt, G.,
\& Yaqoob, T. 2004b, in {\it RAGtime 4/5 Workshops on black holes
and neutron stars}, eds. S. Hled\`{i}k and Z. Stuchl\`{i}k
Opava, (Czech Republic), p. 33 (astro-ph/0407330)

\bibitem[Dov{\v c}iak, Karas, \& Yaqoob 2004]{dovc2004}
Dov\v{c}iak, M., Karas, V., \& Yaqoob, T.\ 2004a, ApJS, 153, 205


\bibitem[Fabian 2004]{fabian2004}
Fabian, A. C., Miniutti, G., Gallo, L., Boller, Th., Tanaka, Y.,
Vaughan, S., \& Ross, R. R. 2004, MNRAS, 353, 1071

\bibitem[Fabian 2005]{fabian2005}
Fabian, A. C., Miniutti, G., Iwasawa, K., \& Ross, R. R. 2005,
MNRAS, 361, 795

\bibitem[Fabian 1989]{fabian1989}
Fabian, A.C., Rees, M.J., Stella, L., \& White, N.E. 1989, MNRAS, 238, 729



\bibitem[Fabian 2002]{fabian2002}
Fabian, A. C. \etal 2002, MNRAS, 335, L1

\bibitem[Fabian \& Vaughan 2003]{fabi2003}
Fabian C., \& Vaughan S. 2003, MNRAS, 340, L28



\bibitem[Gilli \etal 2000]{gilli2000}
Gilli, R., Maiolino, R., Marconi, A., Risaliti, G., Dadina, M.,
Weaver, K. A., \& Colbert, E. J. M. 2000, A\&A, 355, 485

\bibitem[Guainazzi2007]{guain2007}
Guainazzi, M., Bianchi, S. 2007, MNRAS, 374. 1290

\bibitem[Guainazzi2006]{guain2006}
Guainazzi, M., Bianchi, S., \& Dov\v{c}iak, M. 2006, AN, 327, 1032





\bibitem[Kallman 2004]{kallman2004}
Kallman, T. R., Palmeri, P., Bautista, M. A., Mendoza, C., \& Krolik, J. H. 2004, ApJ, 155, 675


\bibitem[magdziarz1995]{magdziarz1995}
Magdziarz, P., \& Zdziarski, A. A. 1995, MNRAS, 273, 837

\bibitem[Matt \etal 2003]{matt2003} 
Matt, G., Guainazzi, M., \& Maiolino, R. 2003 MNRAS, 342, 422


\bibitem[Markowitz 2004]{markowitz2004}
Markowitz, A., \& Edelson, R. 2004, ApJ, 617, 939


\bibitem[Miller 2007]{mill2007} 
Miller, J. M. 2007, ARA\&A, 45, 441

\bibitem[Miller \etal 2008]{miller2008} 
Miller, L., Turner, T. J., \& Reeves, J. N. 2008, A\&A, 483, 437

\bibitem[Miller \etal 2009]{miller2009}
Miller, L., Turner, T. J., \& Reeves, J. N. 2009, MNRAS, in press (astro-ph/0907.3114)

\bibitem[Miniutti et al. 2004]{miniutti2004}
Miniutti, G., \& Fabian, A. C. 2004, MNRAS, 349, 1435

\bibitem[Morse 1995]{morse1995}
Morse, J. A., Wilson, A. S., Elvis, M., \& Weaver, K. A. 1995, ApJ, 439, 121

\bibitem[murphy2007]{murphy2007}
Murphy, K., Yaqoob, T., \& Terashima, Y. 2007, ApJ, 666, 96



\bibitem[Nandra 1997]{nandra1997}
Nandra, K., George, I. M., Mushotzky, R. F., Turner, T. J., \& Yaqoob, T. 1997, ApJ, 488, L91

\bibitem[Nandra 2007]{nandra2007} 
Nandra, K., O'Neill, P. M., George, I. M., \& Reeves, J. N. 2007, MNRAS, 382, 194

\bibitem[Nandra \& Pounds 1994]{nand1994}
Nandra, K., \& Pounds, K. A. 1994, MNRAS, 268, 405



 
\bibitem[Piccinotti \etal 1982]{picc1982}
Piccinotti, G., Mushotzky, R. F., Boldt, E. A., Holt, S. S., Marshall, F. E., 
Serlemitsos, P. J., \& Shafer, R. A. 1982, ApJ, 253, 485

\bibitem[Pike 1996]{pike1996}
Pike, C. D., \etal 1996, ApJ, 464, 487

\bibitem[Ponti 2006]{ponti2006}
Ponti, G., Miniutti, G., Cappi, M.,
Maraschi, L., Fabian, A. C., \& Iwasawa, K. 2006a, MNRAS, 368, 903

\bibitem[Ponti 2006b]{ponti2006b}
Ponti, G., Miniutti, G., Fabian, A. C., Cappi, M., \& Palumbo, 
G. G. C. 2006b, AN, 327, No. 10., 1055


\bibitem[Reynolds 1999]{reynolds1999}
Reynolds, C. S., Young, A. J., Begelman, M. C., \& Fabian, A. C.
1999, ApJ, 521, 99

\bibitem[Rossi2005]{rossi2005}
Rossi, S., Homan, J., Miller, J. M., \& Belloni, T. 2005, MNRAS, 360, 763

\bibitem[Ross \& Fabian 2005]{rossfabian2005}
Ross, R. R., \& Fabian, A. C. 2005, MNRAS, 358, 211





\bibitem[Turner \& Miller 2009]{turnmiller2009}
Turner, T. J., \& Miller, L. 2009, A\&ARv, 17, 47

\bibitem[Turneretal2009]{turneretal2009}
Turner, T. J., \& Miller, L., Kraemer, S. B., Reeves, J. N., \& Pounds, K. A. 
2009, ApJ, 698, 99

\bibitem[Turner \& Pounds 1989]{turn1989}
Turner, T. J., \& Pounds, K. A. 1989, MNRAS, 240, 833

\bibitem[Turner \etal 1991]{turn1991} 
Turner, T. J., Weaver, K. A., Mushotzky, R. F., Holt, S. S.,
 \& Madejski, G. M. 1991, ApJ, 381, 85

\bibitem[Turner \etal 1999]{turner1999}
Turner, T. J., George, I. M., Nandra, K., \& Turcan, D. 1999, ApJ, 524, 667



\bibitem[Weaver \etal 1996]{weav1996} 
Weaver, K. A., Nousek., J., Yaqoob, T., Mushotzky, R. F., Makino, F.,
 \& Otani, C. 1996, ApJ, 458, 160 




\bibitem[Yaqoob \etal 2007]{yaq2007}
Yaqoob, T., Murphy, K. D., Griffiths, R. E. \etal 2007, PASJ, 59, 283 

\bibitem[Yaqoob 2004]{yaqoob2004}
Yaqoob, T., \& Padmanabhan, U. 2004, ApJ, 604, 63

\end{thebibliography}
\end{document}